\documentclass[12pt]{iopart}

\usepackage{tikz}
\usepackage{graphicx}
\usepackage{amssymb}
\usepackage[mathscr]{eucal}

\newcommand{\affiliation}[1]{\address{#1}}

\begin{document}

\title{Brane Bremsstrahlung in DBI Inflation}

\author{Philippe Brax\footnote{philippe.brax@cea.fr} and Emeline Cluzel\footnote{emeline.cluzel@cea.fr}}

\affiliation{Institut de Physique
Th\'eorique, CEA, IPhT, CNRS, URA 2306, F-91191Gif/Yvette Cedex, France  }

\date{today}

\begin{abstract}
We consider the effect of trapped branes on the evolution of a test brane whose motion generates DBI inflation along a warped throat.
The coupling between the inflationary brane and a trapped brane leads to the radiation of non-thermal particles on the trapped brane.
We calculate the Gaussian spectrum of the radiated particles and their backreaction on the DBI motion of the inflationary brane.
Radiation occurs 
either due to a parametric resonance when the interaction time is small compared to the Hubble time or a tachyonic resonance when the interaction time is large.
In both cases the motion of the inflationary brane after the interaction is governed by a chameleonic potential,
which tends to slow it down.
We find that a single  trapped brane can hardly slow down a DBI inflaton whose fluctuations lead to the Cosmic Microwave Background spectrum. A more drastic effect is obtained when the DBI brane encounters a tightly spaced  stack of trapped branes.
\end{abstract}


\section{Introduction}
\label{sec:introduction}
D-brane inflation \cite{Dvali:1998pa, Dvali:2001fw, Burgess:2001fx, Alexander:2001ks, Brodie:2003qv, KKLMMT, Baumann:2009ds} has become a time honoured subject whose development has led to plausible tests of string theory
in a cosmological setting. With the launch of the Planck satellite \cite{planck} and the advent of precision cosmology, this possibility has become even more relevant \cite{Lorenz1, Lorenz2, Lorenz3}. Along these lines, there are three typical observables which may have an impact on models of inflation inspired or derived from string theory. The first one is the spectral index. The second one is the production of gravitational waves \cite{GWS, RK}. Another one is the possibility of primordial non-Gaussianities \cite{Alishahiha:2004eh}. It is quite likely that Planck will reduce the uncertainty on the spectral index in such a way that different scenarios such as brane inflation or modular inflation \cite{Conlon:2005jm, BlancoPillado:2009nw, Barnaby:2009wr} may be distinguished. For instance, modular inflation models of the race track type tend to favour a rather low spectral index \cite{BlancoPillado:2004ns, BlancoPillado:2006he, Brax:2007fe, Brax:2007fz}.
Similarly, the absence of detectable gravity waves in these low field inflation models is generic. In this paper we will consider a related family of brane inflation models which do not fall within the
category of slow roll models. In DBI inflation \cite{Silverstein:2003hf}, a test brane evolves  due to the steepness of the warp factor. This is particularly well motivated in string compactification scenarios where a D3 brane evolves down  a warped throat \cite{KKLT, KS} attracted towards  an anti D3 brane. In general, potential terms such as a large  mass prevent the existence of a slow roll period \cite{KKLMMT}. The brane may thus enter a regime of DBI inflation whose features are very different from the ones of slow roll inflation. In DBI inflation, the spectral index tends to be close to one \cite{Dvali:2001fw, Alishahiha:2004eh, Easson:2009wc} and non-gaussianities are large.
Reheating in brane inflation \cite{Brodie:2003qv, Barnaby:2004gg, Lachapelle:2008sy, Davis:2009wg} is also quite different from reheating in standard inflation \cite{prems, KLS, Felder:1998vq}.

Recently, the issue of brane inflation in the $D3-\bar{D3}$ system has received a host of new developments. It has been realised that the potential does not only comprise the Coulomb interaction and a mass term responsible for the $\eta$ problem \cite{KKLMMT}
; there are also corrections coming from bulk effects in the compactification scheme
\cite{Baumann:2009ni, Baumann:2008kq, Baumann:2006th}.
These bulk effects may require  a detailed knowledge of the full compactification manifold. It happens that the leading contributions may have two effects. One possibility is the existence of an inflection point \cite{Easson:2009kk, Baumann:2007np}
 in the inflation potential when a fractional power of the inflaton in $-c\phi^{3/2}$ is present. Slow roll inflation is tunable in this context and requires a choice of initial conditions to overcome the overshooting problem \cite{overshoot1, overshoot2}.  Another possibility is that the bulk term contributes as another mass term to the potential. In this second case, DBI inflation may  occur.

Our paper discusses brane models without attempting to realise their embedding in string theory. Hints along these lines can be found in \cite{Green:2009ds}.
In this paper, we will focus on DBI inflation as a field theoretical model corresponding to a warped throat with a quadratic potential. We will envisage the likely situation wherein the warped throat is not empty but contains a series of trapped branes \cite{Kofman:2004yc}. These branes may be present due to fixed points in the compactification manifold.
When the inflationary brane crosses these trapped branes, particles are created with a spectrum whose  shape is Gaussian with a width related to the speed of the brane. The radiation of particles leads to the creation of a long range potential which might slow the motion down.
The overall effect of the trapped  brane is to induce a chameleonic term  in the inflationary potential. Eventually this change of the potential implies that the Hubble rate is modified after the crossing of the stack. This effect decays rapidly as the radiated particles are diluted. As a result, the change of the Hubble rate is only effective on a few e-foldings after the interaction.


In this paper we concentrate on the detailed analysis of the creation of particles on the trapped branes and the subsequent slowing down of the inflationary brane in the DBI context.
In section 2, we recall some of the salient characteristics of DBI inflation. In section 3, we consider the creation of particles on the trapped brane. In section 4, we study the backreaction of radiated particules on the inflationary brane and the induced change in the potential.

\section{DBI Inflation}

We are interested in brane models inspired from string theory. The main ingredients of our model will be the DBI nature of the inflaton dynamics and the quartic coupling of the inflaton $\phi$ to a scalar field $\chi$. The coupling is equivalent to the one between the inflaton and brane degrees of freedom on a trapped brane fixed along the inflationary valley \cite{Kofman:2004yc}. In the slow roll regime, the same model has been considered in \cite{Green:2009ds}. Here we analyse the case where the $\eta$ problem prevents slow roll inflation but allows DBI inflation.

The dynamics of the inflationary brane are described by the Dirac-Born-Infeld (DBI) action, where  canonical kinetic terms have been replaced by a DBI term which can be expanded as a sum of higher order derivatives
\begin{eqnarray}
\nonumber
S=-\frac{1}{g_{s}} \int {\rm{d}}^4 x \sqrt {-g} \mbox{ }
\left( \frac{1}{f(\phi)}\sqrt{1+ f(\phi) g_{\mu \nu}\partial^{\mu}\phi \partial^{\nu}\phi} -\frac{1}{f(\phi)}
+g_{\mu \nu}\partial^{\mu}\chi \partial^{\nu}\chi \right.
\\
\label{action}
\left.
+V(\phi)
+ \frac{g^{2}}{2}\chi^{2}|\phi-\phi_{1}|^{2} \right)
+\int {\rm{d}}^4 x \sqrt {-g} \frac{M_{P}^{2}}{2}R
\end{eqnarray}
where $R$ in the Einstein-Hilbert action is the Ricci scalar and $g_{\mu \nu}$ is the metric.
The metric is of the FRW type with no curvature 
\begin{equation}
{\rm{d}}s^2=-{\rm d}t^2+ a^2(t) {\rm d}x^2
\end{equation}
in cosmic time. Conformal time is defined by $a(t) {\rm{d}}\eta= \rm{d}t$.
We will exclusively focus on a warped case with an AdS throat where
 $f(\phi)=\frac{\lambda}{\phi^{4}}$. The 't Hooft coupling $\lambda=R^{4}/\alpha'^{2}=R^{4}/l_{s}^{4}$ depends on the radius of compactification $R$.
 In this scheme, the inflaton represents the radial distance between the moving D3 brane and the $\bar{D3}$ inside the throat : $\phi=\sqrt{T_{3}}r$ where $T_{3}$ is the brane tension.
The coupling constants $g_{s}$ and $g$ are respectively the string coupling and the Yukawa coupling where $g_{s}\approx g^{2}$.

The potential $V(\phi)$ consists of several terms. The Coulomb potential $V_{\rm{D3-\bar{D3}}}=D\left(1-\frac{3D}{16\pi^{2}\phi^{4}}\right)$ with $D=\frac{2}{f(\phi_{\rm{IR}})}$, where $f$ is evaluated at the tip of the throat, corresponds to the attraction between the $D3$ and $\bar{D3}$ branes. There  is also a  mass term  $V_{2}\phi^{2}$ coming from radiative and supergravity corrections to the potential. In general the mass term
$V_{2}$ is  positive apart from  the case of a probe brane starting at the tip of the throat and moving towards the bulk \cite{Chen:2005ad, Kecskemeti:2006cg}.
Corrections to the potential coming from bulk effects have to be added too. These corrections have integer and half-integer powers of $\phi$ and depend on the bulk of the compactification. The two leading  corrections are proportional to $\phi^2$ and $\phi^{3/2}$. In the first case, the model is inflationary when the branes are sufficiently  apart under the influence of the quadratic potential. In the second case, the mass term and a negative $\phi^{3/2}$ term imply the existence of an inflection point around which the potential becomes $ V_0 + V_1 (\phi-\phi_{\rm inflection})$.

There is also a trapped brane whose location is fixed at $\phi_1$ along the throat. Particles $\chi$ on the trapped brane are coupled to the inflationary brane with a quartic coupling at leading order. We will see that this coupling is responsible for the slowing down  of the inflationary brane when crossing the trapped brane.

We define the equivalent of the Lorentz factor  $\gamma$
\begin{equation}
\label{defgamma}
\gamma = \frac{1}{\sqrt{1+ f(\phi) g_{\mu \nu}\partial^{\mu}\phi \partial^{\nu}\phi}}
=\frac{1}{\sqrt{1-f\dot{\phi^{2}}}}.
\end{equation}
The inflationary dynamics are governed by the Klein Gordon equation for the spatially homogeneous inflaton  field $\phi$
\begin{equation}
\label{KGt}
\ddot{\phi}+3H\dot{\phi}+\frac{\dot{\gamma}}{\gamma}\dot{\phi}+\frac{1}{\gamma}\frac{d V}{d \phi}+\frac{1}{\gamma f^{2}}\frac{df}{d\phi}-\frac{1}{\gamma^{2} f^{2}}\frac{df}{d\phi}-\frac{\dot{\phi}^{2}}{2f}\frac{df}{d\phi}=0
\end{equation}
or equivalently
\begin{equation}
\label{KGtdevelop}
\ddot\phi+\frac{3H}{\gamma^{2}}\dot\phi+\frac{1}{\gamma^{3}}\left(\frac{dV}{d\phi}+\frac{1}{f^{2}}\frac{df}{d\phi}\right)-\frac{1}{f^{2}}\frac{df}{d\phi}+\frac{3}{2}\dot\phi^{2}\frac{1}{f}\frac{df}{d\phi}=0
\end{equation}
In the ultrarelativistic limit, both the potential term and the friction term coming from the expansion of the universe are subdominant.

With the warped factor $f(\phi)=\frac{\lambda}{\phi^{4}}$,
equation (\ref{KGt}) becomes :
\begin{equation}
\ddot{\phi}+3H\dot{\phi}+\frac{\dot{\gamma}}{\gamma}\dot{\phi}+\frac{1}{\gamma}\frac{d V}{d \phi}
-\frac{4\phi^{3}}{\lambda\gamma}
+\frac{4\phi^{3}}{\lambda\gamma^{2}}+\frac{2\dot{\phi}^{2}}{\phi}=0
\end{equation}
It is easier to analyse the DBI dynamics using the Hamilton-Jacobi formalism.
This
formalism consists in eliminating the dependence in $t$ and using instead $\phi$-dependent functions.
It is valid provided $\phi(t)$ does not oscillate. 
Using
the Friedmann equation
\begin{equation}
\label{Fried}
H^{2}=\frac{\rho}{3 M_{P}^{2}}
=\frac{1}{3 M_{P}^{2}g_{s}} \left(\frac{\gamma -1}{f}+V \right)
\end{equation}
where $M_{P}$ is the Planck mass, 
the potential can be written as
\begin{equation}
\label{pot}
V(\phi)=3M_{P}^{2}g_{s}H^{2}-\frac{\gamma}{f} + \frac{1}{f}
\end{equation}
If we differentiate (\ref{Fried}) and upon  using the  Klein-Gordon equation  we can express $\frac{dV}{d\phi}$ and obtain
\begin{equation}
\label{point}
\dot{\phi}=-2M_{P}^{2}g_{s}\frac{1}{\gamma}\frac{dH}{d\phi}
\end{equation}
Similarly, the  $\gamma$ factor can be written in terms of functions of $\phi$
\begin{equation}
\label{gamm}
\gamma(\phi) =
\sqrt{1+4M_{P}^{4}g_{s}^{2}f(\phi)\left(\frac{dH}{d\phi}\right)^{2}}
\end{equation}
For a given potential, the Hubble function of $\phi$ is the solution of a differential equation.
Focusing  on the limiting case where $\phi\to 0$ as $t\to \infty$
and expanding  $H(\phi)$ in powers of $\phi$, the relativistic factor is dominated by
\begin{equation}
\gamma(\phi) \approx
2M_{P}^{2}g_{s}\sqrt{f(\phi)}\left|\frac{dH}{d\phi}\right|
\end{equation}
As a result, the inflaton follows a universal trajectory dictated by the coupling $f(\phi)$
\begin{equation}
\dot \phi \approx -\frac{1}{\sqrt{ f(\phi)}}
\quad
\mbox{       and       }
\quad
\phi(t)=\frac{\sqrt{\lambda}}{t}
\label{evol}
\end{equation}
The potential term has an influence on the scale factor only.

We will exclusively focus on a quadratic potential where
\begin{equation}
V(\phi)=m^2\phi^2
\end{equation}
In this case,
there are co-dominant terms in (\ref{pot}) and one finds
\begin{equation}
H=\frac{1}{3\sqrt{\lambda}} \left( 1+\sqrt{1+\frac{3m^2\lambda}{M_{P}^{2}g_{s}}} \right) \phi
\end{equation}
The scale factor becomes
\begin{equation}
\frac{a}{a_{0}}=\left(\frac{t}{t_{0}}\right)^{1/\epsilon}
\end{equation}
where
\begin{equation}
\frac{1}{\epsilon} = \frac{1}{3} \left( 1+\sqrt{1+\frac{3m^2\lambda}{M_{P}^{2}g_{s}}} \right)
\approx \sqrt{\frac{\lambda}{3g_{s}}}\frac{m}{M_{P}}.
\end{equation}
Inflation occurs when $\epsilon < 1$. We will focus on the case where $\epsilon \ll 1$ which leads to interesting restrictions on the parameter space (see section 4). In all cases the coupling
 $\frac{\lambda}{g_{s}}$ and $\frac{m}{M_{P}}$ can be adjusted to verify $\epsilon\ll 1$ at  strong coupling $\lambda \gg 1$.
The $\gamma$ factor becomes
\begin{equation}
\gamma
=\frac{2M_{P}^{2}g_{s}}{\lambda}\frac{1}{\epsilon}t^{2}
\end{equation}
In the following we shall use conformal time to study the creation of particles when the DBI brane crosses the trapped brane.
In conformal time we find that
\begin{equation}
\label{aQconf}
 a(\eta)\propto \left(\frac{1-\epsilon}{\epsilon}\right)^{\frac{1}{\epsilon-1}} (\frac{\eta}{\eta_0})^{\frac{1}{\epsilon-1}}
\end{equation}
in such a way that
\begin{equation}
\label{HQconf}
{\cal{H}} \approx -\eta^{-1}, \  \  \
\frac{a''}{a} \approx 2\eta^{-2}
\end{equation}
We deduce $\phi\propto (\frac{\eta}{\eta_0})^{-\frac{\epsilon}{\epsilon-1}}\approx (\frac{\eta}{\eta_0})^{\epsilon}$ to leading order.
We will make use of these identities in conformal time when discussing particle creation.

\section{Particle Creation}

\label{sec:WKB}
\subsection{WKB approximation}
So far we have not taken into account the presence of the trapped brane. In fact, the trapped brane has an influence on the
inflationary brane evolution. To determine its effect we need to study the quantum modes of the field $\chi$.
Let us expand the quantum field $\chi$ in terms of creation and annihilation operators. For convenience sake we work in conformal time. The field reads
\begin{equation}
\chi(\eta)=\int \frac{{\rm d}^3 k}{(2\pi)^3}(a_{k}\chi_{k}(\eta)e^{ikx}+a_{k}^{\dagger}\chi_{k}^{*}(\eta)e^{-ikx})
\end{equation}
and each mode satisfies the Klein-Gordon equation
\begin{equation}
\chi''_{k}+ 2 {\cal H} \chi'_{k}+ k^{2}\chi_{k}+a^{2}g^{2}|\phi-\phi_{1}|^{2}\chi_{k}=0
\end{equation}
We can put in a Schrodinger form by defining
\begin{equation}
\Psi(\eta)={a}\chi(\eta)
\end{equation}
So $\Psi$  obeys
\begin{equation}
\label{bas}
\Psi_{k}''+ \omega_{k}^{2}\Psi_{k}=0
\end{equation}
with a time dependent frequency
\begin{eqnarray}
\omega_{k}(\eta)=\sqrt{k^{2}+A(\eta)},\ \ \
A(\eta)=a^{2}g^{2}|\phi-\phi_{1}|^{2}-\frac{a''}{a}
\end{eqnarray}
where $\phi$ is the unperturbed $\phi(\eta)$ corresponding to the unperturbed motion of the inflationary brane.
Here we have assumed $\omega_{k}^{2}$ positive. When it is negative, the regime is tachyonic and the frequency will be written $\Omega_{k}=\pm\rm{i}\omega_{k}=\pm\rm{i}\sqrt{|k^{2}+A(\eta)|}$, so that $\Omega_{k}^{2}=-\omega_{k}^{2}>0$.

The equation for the modes (\ref{bas}) can be approximately solved using the Wentzel-Kramers-Brillouin or WKB approximation. Far in the past, the solution is assumed to be in a Bunch-Davies vacuum where
 for $\eta\rightarrow -\infty$ we have
\begin{equation}
\Psi_{k}(\eta)=\frac{1}{\sqrt{2\omega_{k}(\eta)}}\quad e^{-i\int^{\eta}\omega_{k}(\eta')\rm{d}\eta'}
\end{equation}
When the inflationary brane has passed through the trapped brane and
 $\eta\rightarrow 0$, it is a mixture of  two possible modes

\begin{equation}
\Psi_{k}(\eta)=\frac{\alpha_{k}(\eta)}{\sqrt{2\omega_{k}(\eta)}}\quad e^{-i\int^{\eta}\omega_{k}(\eta')\rm{d}\eta'}+\frac{\beta_{k}(\eta)}{\sqrt{2\omega_{k}(\eta)}}\quad e^{i\int^{\eta}\omega_{k}(\eta')\rm{d}\eta'}
\end{equation}
where $\alpha_{k}$ and $\beta_{k}$ are the Bogoliubov coefficients.
The WKB approximation is valid when
$|\frac{{\omega}'}{\omega^{2}}|\ll1$.
We define
\begin{equation}
R=\left|\frac{{\omega}'}{\omega^{2}}\right|
\end{equation}
and find
\begin{equation}
R=\left|\frac{{\cal{H}}g^{2}a^{2}|\phi-\phi_{1}|^{2}+\phi'g^{2}a^{2}|\phi-\phi_{1}|
-\frac{1}{2}\left(\frac{a''}{a}\right)'}{(k^{2}+a^{2}g^{2}|\phi-\phi_{1}|^{2}-\frac{a''}{a})^{3/2}}\right|
\end{equation}

Our goal is to determine under which conditions the WKB approximation is violated. There are two physically different situations where the analysis can be easily carried out. They depend on
\begin{equation}
\xi= \frac{H^2}{g|\dot\phi|}
\end{equation}
which is a constant in DBI inflation with a quadratic potential
\begin{equation}
\xi= \frac{1}{g\epsilon^2 \sqrt \lambda}\approx \frac{\sqrt{\lambda}}{3g^3}\left(\frac{m}{M_P}\right)^2
\end{equation}
The creation of particles is very different for small or large $\xi$.

\subsection{Small $\xi$ behaviour}

In this region of the parameter space, the creation of particles occurs when the DBI brane is close to the trapped brane which corresponds to the regime where
\begin{equation}
\vert \phi -\phi_1\vert \ll \frac{|\dot \phi|}{H}
\end{equation}
In this case
the $R$ factor becomes
\begin{equation}
\label{near}
R \approx R_{\rm{near}} = \frac{|\phi' g^{2}a^{2}|\phi-\phi_{1}|-\frac{1}{2}\left(\frac{a''}{a}\right)'|}
{|k^{2}+a^{2}g^{2}|\phi-\phi_{1}|^{2}-\frac{a''}{a}|^{3/2}}
\end{equation}
and   non-adiabaticity is mainly present  in the region where $g^2 a^2 \vert \phi-\phi_1\vert^2 \gg \frac{a''}{a}$ (or equivalently $g^2 \vert \phi-\phi_1\vert^2 \gg 2H^2$) for which we have
\begin{equation}
R \approx \left|\frac{\phi'g^{2}a^{2}|\phi-\phi_{1}|
}{(k^{2}+a^{2}g^{2}|\phi-\phi_{1}|^{2})^{3/2}}\right|
\end{equation}
From now on
we define
\begin{equation}
{\cal{K}}=\frac{k}{a}
\end{equation}
which corresponds to physical momenta,

The creation of particles arises when $R>1$ at its maximum which is located
at
$g^2\vert \phi-\phi_1\vert^2 =\frac{{\cal{K}}^2}{2}$
and is given by
\begin{equation}
R_{\rm max}= \frac{2}{3^{3/2}} \frac{ g |\dot\phi|}{{\cal{K}}^2}
\end{equation}
implying that the creation of particles occurs when ${\cal{K}} \le \frac{\sqrt 2}{3^{3/4}} \sqrt {g|\dot \phi|}$ . The maximal extension of the non-adiabatic region is
\begin{equation}
\Delta\phi  = \frac{1}{3^{3/4}} \sqrt \frac{|\dot \phi|}{g}
\end{equation}
Notice that $\Delta \phi \gg \frac{H}{g}$ as $\xi \ll 1$.

Inside a  region of size $\vert \phi-\phi_1 \vert \le H/g$ around the origin, there is a domain of non-adiabaticity when
\begin{equation}
R \approx \frac{|\left(\frac{a''}{a}\right)'|}
{2|k^{2}-\frac{a''}{a}|^{3/2}}
>1
\end{equation}
corresponding to
\begin{equation}
\label{physco}
\sqrt{2-2^{2/3}} H \le {\cal{K}} \le \sqrt{2+2^{2/3}} H
\end{equation}
Due to the very small width of this zone in momentum space, virtually no particles are created in this interval.

Finally  when ${\cal{K}} \le \sqrt{2} H$, there is a tachyonic instability. The size of the tachyonic region is much smaller than the size of the non-adiabaticity region where most particles are created. In fact, the creation of particles in the tachyonic region is negligible as it scales like
$\exp(H\Delta t_{\rm tachyon})\sim \exp (\xi) \sim 1$ where the time spent in the tachyonic region is $\Delta t_{\rm tachyon} \sim \frac{H}{g|\dot \phi|}$.

It is interesting to notice that 
the time spent by the brane in  
the interaction region is
\begin{equation}
\Delta t\approx  \frac{\sqrt \xi}{H}
\end{equation}
implying that the interaction lasts less than a Hubble time and therefore the interaction time corresponds to
a number of e-folds
\begin{equation}
\Delta N= \frac{\Delta a}{a} \approx \sqrt{\xi}\ll 1
\end{equation}
The interaction is almost instantaneous.

So far we have used the fact that the Hubble rate is nearly constant in the interaction region.
This can be ascertained  as the variation of the Hubble rate in the interaction region is given by
\begin{equation}
\left|\frac{\Delta{{H}}}{{H}}\right| \approx \frac {1}{(g\sqrt\lambda)^{1/2}}
\end{equation}
Therefore we must impose that  $ g\sqrt \lambda \gg 1$. We shall see that this is always the case when $\epsilon \ll 1$.

\subsection{Large $\xi$ behaviour}

In this case, the ratio $R$ can be simplified in two regimes  depending on whether the DBI brane has moved far from the trapped brane  or not.
If the inflationary  brane is far from the trapped brane  then $|\phi-\phi_{1}|\gg \dot\phi / {H}$ and
\begin{equation}
R \approx R_{\rm{far}} = \frac{|{\cal{H}}g^{2}a^{2}|\phi-\phi_{1}|^{2}-\frac{1}{2}\left(\frac{a''}{a}\right)'|}
{|k^{2}+a^{2}g^{2}|\phi-\phi_{1}|^{2}-\frac{a''}{a}|^{3/2}}
\end{equation}
On the contrary, if they are close, $|\phi-\phi_{1}|\ll \dot\phi / {H}$ so
\begin{equation}
\label{near}
R \approx R_{\rm{near}} = \frac{|\phi' g^{2}a^{2}|\phi-\phi_{1}|-\frac{1}{2}\left(\frac{a''}{a}\right)'|}
{|k^{2}+a^{2}g^{2}|\phi-\phi_{1}|^{2}-\frac{a''}{a}|^{3/2}}
\approx \frac{|\left(\frac{a''}{a}\right)'|}
{2|k^{2}-\frac{a''}{a}|^{3/2}}
\end{equation}
These two regimes capture all the physics of the $\chi$-particle creation.

First, we will study the case when the two branes are far from each other.
The ratio $R$ is maximal when $g^{2}|\phi-\phi_{1}|^{2}=2{\cal{K}}^{2}+(2-7\epsilon)H^2$ and its value is simply
\begin{equation}
R_{\rm max}
= \frac{|2H{\cal{K}}^{2}-4\epsilon H^{3}|}{|3{\cal{K}}^{2}-6\epsilon H^{2}|^{3/2}}
\end{equation}
We find that there is  a pole  for ${\cal{K}}_{\rm{pole}}=\sqrt{2\epsilon} H$
. The value at the origin is very large: $R_{\rm max}({\cal{K}}=0)\approx \epsilon^{-1/2}\gg 1$ as $\epsilon\ll 1$. In all the interval between the origin and the pole, $R_{\rm max}$ is greater than one. We  want to determine the physical momentum ${\cal{K}}_{\rm{max}}$ for which $R_{\rm max}=1$ and then becomes smaller than unity for larger momenta. Expanding around the pole
\begin{equation}
R_{\rm max}=\frac{2}{3\sqrt{3}} \frac{H}{({\cal{K}}^2-{\cal{K}}^2_{\rm{pole}})^{1/2}}
\end{equation}
implying that
\begin{equation}
{\cal{K}}_{\rm{max}}^{2} \approx \frac{4H^{2}}{27}.
\end{equation}
Hence we find that for physical momenta $0<{\cal{K}}<{\cal{K}}_{\rm{max}}\approx \frac{2{H}}{3\sqrt{3}}$, the WKB approximation is  violated around $g^{2}|\phi-\phi_{1}|^{2}=2({\cal{K}}^2+H^2)$.
Therefore there are two non-adiabatic regions far from the trapped brane: regions I and II centered respectively around  $\phi_{A}$ and $\phi_{B}$.
The approximation used here is valid as the maximal extension of the interaction zone is given by $\Delta \phi \approx \frac{H}{g}$ implying that
$\Delta \phi \gg |\dot\phi| /H$ when $\xi\gg 1$.

We now consider  when the inflationary brane  and the trapped  brane are close to each other.
From (\ref{near}), we find that $R>1$ when
\begin{equation}
\label{ineg}
\sqrt{2-2^{2/3}}H<{\cal{K}}<\sqrt{2+2^{2/3}}H
\end{equation}
The inequality (\ref{ineg}) gives the range of physical momenta for which a non-adiabatic region appears in the immediate vicinity of the trapped brane. It is consistent with   $g^2 \vert \phi-\phi_1\vert^2 \ll 2H^2$ as $\xi \gg 1$.

On top of the non-adiabatic instability detailed above, there is a tachyonic resonance when $\omega_{k}^{2}<0$ corresponding to
\begin{equation}
\label{tachyon}
{\cal{K}}^{2}+g^{2}|\phi-\phi_{1}|^{2}-2H^2<0
\end{equation}
So for physical momenta larger than $2H^2$, there is no tachyonic regime.
We define $\eta_{-}$ and $\eta_{+}$ as the two turning points such that $\omega_{k}^{2}(\eta_{-})=\omega_{k}^{2}(\eta_{+})=0$.
There is a physical momentum ${\cal{K}}^*$ for which the non-adiabatic and tachyonic  regions intersect in just one point. For $0<{\cal{K}}<{\cal{K}}^*$, the regions intersect and for ${\cal{K}}^*<{\cal{K}}<{\cal{K}}_{\rm{max}}$ the tachyonic region and the non-adiabatic region are disconnected.
In the tachyonic regime, we also use the WKB approximation. It is valid when $\left|\frac{\Omega_{k}'}{\Omega_{k}^{2}}\right|<1$ and the modes are
\begin{equation}
\Psi_{k}(\eta)=\frac{a_{k}(\eta)}{\sqrt{2\Omega_{k}(\eta)}}\quad e^{-\int^{\eta}\Omega_{k}(\eta')\rm{d}\eta'}+\frac{b_{k}(\eta)}{\sqrt{2\Omega_{k}(\eta)}}\quad e^{\int^{\eta}\Omega_{k}(\eta')\rm{d}\eta'}
\end{equation}
The Bogoliubov coefficients change after going through a non-adiabatic region.

The interaction time can be estimated and gives
\begin{equation}
H\Delta t \approx {\xi}
\end{equation}
leading to
\begin{equation}
\Delta N= \frac{\Delta a}{a}\approx \xi
\end{equation}
So the interaction region is spread out over a large number of efoldings.
Moreover we must impose that the variation of the Hubble rate is small in the interaction region
\begin{equation}
\frac{\Delta H}{H}\approx \frac{1}{g\epsilon \sqrt \lambda}
\end{equation}
hence we must have $g\epsilon \sqrt \lambda \gg 1$.

\subsection{Creation of particles}

We are interested in the particles created when the inflationary brane crosses the trapped brane. This happens when the WKB approximation breaks down.
Let us first concentrate on the $\xi\ll 1$ regime.
The calculation of the number of particles in this region is well-known \cite{Kofman:2004yc}.
The Bogoliubov coefficient is obtained by expanding
\begin{equation}
\frac{\omega_{k}}{a}
\approx
\sqrt{{\cal{K}}^2+g^{2}|\phi-\phi_{1}|^{2}}\approx g \vert \dot\phi_1\vert \delta t \left(1- \frac{{\cal{K}}^2}{2g^2\dot \phi_1^2(\delta t)^2}\right)
\end{equation}
around $\phi_1$ ($\delta t = t-t_{1}$) and integrating around a contour where the WKB approximation is still valid in the complex plane. The end result is that
\begin{equation}
\vert\beta^f_k\vert^2= e^{-\pi \frac{{\cal{K}}_1^2}{g \vert \dot\phi_1\vert }}
\end{equation}
where ${\cal K}_1= k/a_1$.
As a result, the spectrum is Gaussian with a width determined by the speed of the brane.

In the $\xi\gg 1$ regime we have found
different situations depending on the physical momentum ${\cal{K}}$.
The configuration where $0<{\cal{K}}<{\cal{K}}_{\rm{max}}$ is the most complex. There are two symmetric non-adiabatic regions far from $\phi_{1}$ and a tachyonic region in the proximity of $\phi_{1}$.
The tachyonic region intersect with the non-adiabatic regions for $0<{\cal{K}}<{\cal{K}}^{*}<{\cal{K}}_{max}$.
For ${\cal{K}}_{\rm{max}}<{\cal{K}}<\sqrt{2-2^{2/3}}H$, the tachyonic region is still there but we no longer have any non-adiabatic region.
For $\sqrt{2-2^{2/3}}H<{\cal{K}}<\sqrt{2}H $, there is a tachyonic region and inside of it a non-adiabatic zone.
For $\sqrt{2}H<{\cal{K}}<\sqrt{2+2^{2/3}}H$, there is no tachyonic resonance but there is a non-adiabatic region around $\phi_{1}$.
And finally for any ${\cal{K}}>\sqrt{2+2^{2/3}}H$, there is no tachyonic resonance and the regime is always adiabatic. Notice that all these intervals are easy to interpret using the physical wave number $\cal{K}$. The size of the physical intervals in $\cal{K}$ is time-independent.

Let us  first study the creation of particles for $0<{\cal{K}}^{*}<{\cal{K}}<{\cal{K}}_{\rm{max}}$. This case is typical and will allow us to deduce the particle spectrum in the other intervals too.

\vskip 2 cm
\begin{figure} [htbp]

\begin{tikzpicture}

\draw[->](0,0) -- (15,0);
%
\draw[<->][red](0.8,-0.8) -- (3.2,-0.8);
\draw[<->][red](8.8,-0.8) -- (11.2,-0.8);
\draw[blue][<->] (4,1) -- (8,1);
%
\draw[<-] (4.2,-0.1) -- (5,-0.6);
\draw[->] (7,-0.6) -- (7.8,-0.1);
\draw(2,0) node[below,scale = 1]{$\eta_{A}$};
\draw(4,0) node[below,scale = 1]{$\eta_{-}$};
\draw(8,0) node[below,scale = 1]{$\eta_{+}$};
\draw(10,0) node[below,scale = 1]{$\eta_{B}$};
\draw(6,0) node[above,scale = 1]{$\eta_{1}$};
\draw(15,0) node[below,scale = 1]{$\Re(\eta)$};
\draw(0,0) node[below,scale = 1]{$-\infty$};
\draw(2,-1) node[below,scale = 0.7]{non-adiabatic I};
\draw(10,-1) node[below,scale = 0.7]{non-adiabatic II};
\draw(6,-0.5) node[below,scale = 0.7]{turning points};
\draw(6,-0.8) node[below,scale = 0.7]{($\omega^{2}=0$)};
\draw(6,1.1) node[above,scale = 0.7]{tachyonic zone};
%
\foreach \x in {2, 4, 6, 8, 10}
    \draw (\x,-2pt) -- (\x,2pt);
%
\foreach \x in {4.1, 4.4 , 4.7 , 5 , 5.3 , 5.6 , 5.9, 6.2, 6.5 , 6.8, 7.1, 7.4 , 7.7 }
    \draw[blue] (\x-0.1,-2pt) -- (\x+0.1,2pt);
\draw[red,very thick] (0.8,0) arc(180:0:1.2);
\draw[orange,very thick] (3.5,0) arc(180:0:0.5);
\draw[orange,very thick] (7.5,0) arc(180:0:0.5);
\draw[red,very thick] (8.8,0) arc(180:0:1.2);
\end{tikzpicture}
\caption {Configuration of the interaction zone in the complex plane \\ for ${\cal{K}}^{*}<{\cal{K}}<{\cal{K}}_{\rm{max}}$}
\end{figure}
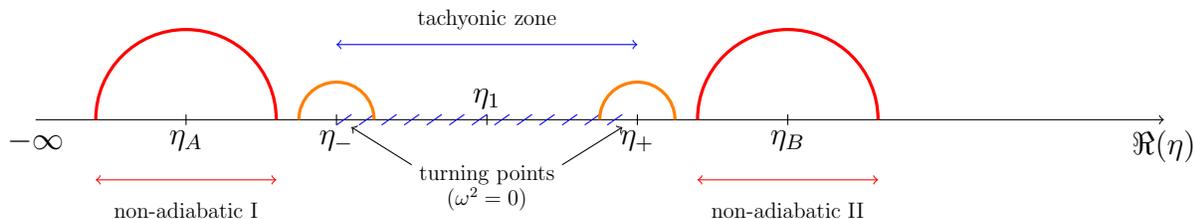
\vskip 2 cm 
Initially, for $\eta \rightarrow -\infty$, the modes are in a Bunch-Davies vacuum.
Then the WKB approximation breaks down in the non-adiabatic region I, far enough from it the modes become
\begin{equation}
\Psi_{k}(\eta)=\frac{\alpha_{k}(\eta)}{\sqrt{2\omega_{k}(\eta)}}\quad e^{-i\int^{\eta}\omega_{k}(\eta')d\eta'}+\frac{\beta_{k}(\eta)}{\sqrt{2\omega_{k}(\eta)}}\quad e^{i\int^{\eta}\omega_{k}(\eta')d\eta'}
\end{equation}
We need to determine the two Bogoliubov coefficients. In fact they can be expressed as  transmission and reflection coefficients $t$ and $r$
\begin{equation}
\label{coefcoef}
\alpha_{k}(\eta<\eta_{A})=t \alpha_{k}(\eta>\eta_{A})
\mbox{    and    }
\beta_{k}(\eta>\eta_{A})=r \alpha_{k}(\eta>\eta_{A})
\end{equation}
where the conformal time $\eta$ increases.
So with the chosen initial condition $\alpha_{k}(-\infty)=1$, this gives
\begin{equation}
\alpha_{k}(\eta>\eta_{A})=1/t, \ \
\beta_{k}(\eta>\eta_{A})=r/t
\end{equation}
The solution before the non-adiabatic region I is linked to the solution on the other side of the non-adiabatic region by an analytic continuation in the complex plane. We consider that the time variable $\eta$ is a complex variable. We draw a contour around $\eta=\eta_{A}$ in the complex plane (see figure 1). The radius of this semi-circle must be large enough for the WKB approximation to be valid before and after the non-adiabatic region. This is the case if
\begin{equation}
\omega_{k}(\eta_{A})|\eta-\eta_{A}|\gtrsim R(\eta_{A})=R_{\rm{max}}
\end{equation}
and we trust the expansion of  $\omega_{k}^{-1}$ around $\eta_{A}$
\begin{equation}
\frac{1}{\omega_{k}(\eta)} = \frac{1}{\omega_{k}(\eta_{A})}-\left(\frac{\omega_{k}'}{\omega_{k}^{2}}\right)(\eta_{A})(\eta-\eta_{A})
-\sum_{n=2}^{\infty}\frac{R_{A}^{(n-1)}}{n!}(\eta-\eta_{A})^{n}
\end{equation}
where $R_{A}^{(n)}$ is the n-th derivative of $R$ at $\eta_{A}$.
The condition on the size of the semi-circle implies that
\begin{equation}
\omega_{k}(\eta) \approx -\frac{1}{R_{\rm{max}}(\eta-\eta_{A})}
\end{equation}
Using the fact that in the non-adiabatic region $ \omega_{k}(\eta_{A})=O(aH)$ and $R_A=O(\frac{H}{{\cal{K}}})=O(1)$, the condition on the contour is equivalent to $\delta t \gtrsim 1/H$. Now the duration of the interaction region is $ \frac{H}{g|\dot \phi|}$ implying that the contour circling around the non-adiabatic region is much smaller than the interaction region.
As a result we obtain that
\begin{equation}
\exp\left(\pm {\rm{i}}\int \omega_{k}(\eta'){\rm d}\eta'\right)
\approx \exp\left(\mp {\rm{i}}\int \frac{1}{R_{\rm{max}}(\eta'-\eta_{A})} {\rm d}\eta'\right)
\end{equation}
Notice that $\eta-\eta_{A}$ is negative before the non-adiabatic region and positive later on. A positive-frequency mode is changed in a negative-frequency mode when going through the non-adiabatic region (and vice versa).
With $(\eta-\eta_{A})=\rho e^{\rm{i}\theta}$ and $\rm{d}\eta=\rm{i}\rho e^{\rm{i}\theta} \rm{d}\theta$, the contour integral is given by the residue theorem.
A factor of $\rm{i}$ also appears from $\eta-\eta_{A}\rightarrow e^{\rm{i}\pi} (\eta_{A}-\eta)$ in $\frac{1}{\sqrt{2\omega_{k}}}$ so finally
\begin{equation}
\label{firstbeta}
\beta_{k}=-\rm{i} \exp(-2\rm{i}\theta^{A}) \exp\left(\frac{\pi}{R_{\rm{max}}}\right)
=-\rm{i} \exp(-2\rm{i}\theta^{A}) \exp\left(\frac{3\sqrt{3}\pi}{2} \frac{{\cal{K}}_A}{H}\right)
\end{equation}
with ${\cal K}_A= k/a_A$ and the phase is
\begin{equation}
\theta^{A}=\int_{-\infty}^{\eta_{A}} \omega_{k} \rm{d}\eta
\end{equation}
Probability conservation  imposes $|\alpha_{k}|^{2}-|\beta_{k}|^{2}=1$, so we deduce
\begin{equation}
\label{firstalpha}
\alpha_{k}=e^{\rm{i}\varphi}\sqrt{1+\exp\left(3\sqrt{3}\pi \frac{{\cal{K}}_A}{H}\right)}
\end{equation}
where $\varphi$ is a random phase.

We now compute the Bogoliubov coefficient $b_{k}$ of the non-vanishing wave in the tachyonic region \cite{Dufaux:2006ee}. We neglect the decaying mode in this region depending on the coefficient $a_k$.
We draw a contour around the turning point $\eta_{-}$ where $\omega_{k}^{2}(\eta_{-})=0$. We assume the radius of the semi-circle is large enough for the WKB aproximation to be valid along the contour  :
\begin{equation}
\omega_{k}^{2}=\frac{d\omega_{k}^{2}}{d\eta}(\eta_{-})(\eta-\eta_{-})
\end{equation}
After an analytic continuation we find
\begin{equation}
b_{k}=\alpha_{k}e^{-\rm{i}(\theta^{-}+\frac{\pi}{4})}+\beta_{k} e^{\rm{i}(\theta^{-}+\frac{\pi}{4})}
\end{equation}
with the phase
\begin{equation}
\theta^{-}=\int_{-\infty}^{\eta_{-}}\omega_{k}\rm{d}\eta
\end{equation}
depending on the wave evolution before the turning point.
Then passing around the second turning point at $\eta_{+}$, we obtain a new contribution to the Bogoliubov coefficients.
\begin{equation}
\label{tildebetak}
\tilde\beta_{k}=e^{\int_{\eta_{-}}^{\eta_{+}}\Omega \rm{d}\eta} e^{-\rm{i}(\theta^{-}+\frac{\pi}{4})} b_{k}
=e^{\int_{\eta_{-}}^{\eta_{+}}\Omega \rm{d}\eta} \left(\beta_{k}+\alpha_{k} e^{-2\rm{i}(\theta^{-}+\frac{\pi}{4})}\right)
\end{equation}
\begin{equation}
\label{tildealphak}
\tilde\alpha_{k}=e^{\int_{\eta_{-}}^{\eta_{+}}\Omega \rm{d}\eta} e^{\rm{i}(\theta^{-}+\frac{\pi}{4})} b_{k}
=e^{\int_{\eta_{-}}^{\eta_{+}}\Omega \rm{d}\eta} \left(\alpha_{k}+\beta_{k} e^{2\rm{i}(\theta^{-}+\frac{\pi}{4})}\right)
\end{equation}
The final contribution comes from the non-adiabatic region II where the wave appears to break into reflected and transmitted ones
\begin{equation}
\label{passage}
\tilde\alpha_{k}\chi_{-} \rightarrow R_{-}\chi_{+}+T_{-}\chi_{-}
\end{equation}
\begin{equation}
\label{dual}
\tilde\beta_{k}\chi_{+} \rightarrow R_{+}\chi_{-}+T_{+}\chi_{+}
\end{equation}
We are particularly interested in the
the final Bogoliubov coefficient
\begin{equation}
\label{betakf}
\beta_{k}^{f}=R_{-}+T_{+}
\end{equation}
The first situation (\ref{passage}) is the same as in (\ref{coefcoef}) but with a phase $\theta^{B}$.
\begin{equation}
\label{Rmoins}
R_{-} = \frac{r}{t} \tilde\alpha_{k}
= -\rm{i} \exp(-2\rm{i}\theta^{B}) \exp\left(\frac{3\sqrt{3}\pi}{2} \frac{{\cal{K}}_B}{H}\right) \tilde\alpha_{k}
\end{equation}
where ${\cal K}_B= k/a_B$
The second situation (\ref{dual}) is the dual configuration for $\Psi^*$,
where we find
\begin{equation}
\label{Up}
|T_{+}|^{2}=|\tilde\beta_{k}|^{2}\left(1+\left|\frac{r^*}{t^*}\right|^{2}\right)
\end{equation}
via current conservation. Therefore we can introduce another phase $\vartheta$ such that
\begin{equation}
\label{2conservation}
T_{+}=e^{\rm{i}\vartheta}\tilde\beta_{k}\sqrt{1+e^{3\sqrt{3}\pi \frac{{\cal{K}}_B}{H}}}
\end{equation}
As a result we  find the Bogoliubov coefficient
\begin{eqnarray}
\nonumber
\beta_{k}^{f}=-\rm{i} \mbox{ }
e^{\int_{\eta_{-}}^{\eta_{+}}\Omega \rm{d}\eta}
\left(
e^{2\rm{i}(\theta^{-}-\theta^{A}-\theta^{B})}e^{3\sqrt{3}\pi \frac{{\cal{K}}_A+{\cal K}_B}{2H}} +
e^{\rm{i}\varphi+\rm{i}\vartheta-2\rm{i}\theta^{-}}
\sqrt{1+e^{3\sqrt{3}\pi\frac{{\cal{K}}_A}{H}}}\sqrt{1+e^{3\sqrt{3}\pi\frac{{\cal{K}}_B}{H}}}
\right.
\\
\label{phasesal}
\left.
+
e^{\rm{i}\varphi-2\rm{i}\theta^{B}}e^{\frac{3\sqrt{3}\pi}{2} \frac{{\cal{K}}_B}{H}} \sqrt{1+e^{3\sqrt{3}\pi
\frac{{\cal{K}}_A}{H}}}+e^{\rm{i}\vartheta-2\rm{i}\theta^{A}}e^{\frac{3\sqrt{3}\pi}{2} \frac{{\cal{K}}_A}{H}} \sqrt{1+e^{3\sqrt{3}\pi
\frac{{\cal{K}}_B}{H}}}
\mbox{ }
\right)
\end{eqnarray}

Let us now examine the other regions.
For ${\cal{K}}_{\rm max}^{2}<{\cal{K}}^{2}<(2-2^{2/3})H^2$, there is a purely tachyonic contribution.
For $\alpha_{k}(-\infty)=1$, we know from (\ref{tildebetak}) that
\begin{equation}
\label{nbretach}
\vert  \beta_{k}^f\vert^2=e^{2\int_{\eta_{-}}^{\eta_{+}}\Omega_{k}d\eta}
\end{equation}
The same behaviour is valid for $(2-2^{2/3})H^2<{\cal{K}}^{2}<2H^2$ because the non-adiabatic region has no importance when it is included in the tachyonic regime. Indeed the passage through the non-adiabatic region in a tachyonic region only changes the phase of the Bogoliubov coefficients.

When $2H^2<{\cal{K}}^2<(2+2^{2/3})H^2$, there is a non-adiabatic region centered at $\phi_{1}$. In this region where ${\cal{K}}$ is large, the duration of the interaction  is less than $1/H$.
We continue the wave function in the complex plane on a semi-circle of radius larger than $1/H$  such that
\begin{equation}
\frac{\omega_{k}}{a}
\approx \sqrt{{\cal{K}}^2-2H^2+ g^2 \dot\phi_1^2 \delta t^2}
\approx g \dot\phi_1 \delta t + \frac{{\cal{K}}^2-2H^2}{2 g \dot\phi_1 \delta t}
\end{equation}
 Notice that the WKB approximation is valid along this circle. The integral $\int  \frac{\omega_k}{a} dt$ picks up an imaginary value due to the residue at the origin.
As in (\ref{firstbeta}) positive and negative-frequency modes are exchanged. And the coefficient $\beta_{k}$ is computed with the residue theorem
\begin{equation}
\vert  \beta_{k}^f\vert^2=e^{\pi \frac{ 2H^2- {{\cal{K}}_1}^2}{g\vert \dot\phi_1\vert}}
\end{equation}
Notice that this is a small number in the region where ${\cal{K}}\ge O(\sqrt 2H)$.

For $0<{\cal{K}}<{\cal{K}}^*$, the method used for ${\cal{K}}^*<{\cal{K}}<{\cal{K}}_{\rm{max}}$ is not reproducible as the semi-circles drawn around the non-adiabatic regions I and II intersect the semi-circles drawn around the turning points. We will see that the low energy part of the spectrum is dominated by the tachyonic instability and therefore does not really depend on the non-adiabatic region.  As a result
we will extend our results to the whole momentum range $0<{\cal{K}}<{\cal{K}}_{\rm{max}}$.

We have now obtained the Bogoliubov coefficients in both regimes for $\xi$.
This is particularly important as one can define
an adiabatic invariant for equation (\ref{bas})
\begin{equation}
{\cal N}_k=\frac{\omega_{k}}{2}\left(\frac{|\Psi_{k}'|^{2}}{\omega_{k}^{2}}+|\Psi_{k}|^{2}\right)-\frac{1}{2}
=|\beta^f_{k}|^{2}
\end{equation}
As explained in \cite{KLS}, this is the comoving occupation number of particles with momentum $k$.

\section{A DBI Chameleon}

\subsection{The modified potential}

We will compute the energy density of the created particles. In fact this energy density appears to add a contribution to the potential of the inflaton. This effect corresponds to the slowing down of the inflationary brane by the trapped brane. The emitted energy density is
\begin{equation}
\rho_\chi= \int\frac{{\rm{d}}^{3}{\cal{K}}}{(2\pi)^{3}} \mbox{ } \tilde\omega_{k} \mbox{ } {\cal N}_{k}
\end{equation}
where the integration is done over the physical momenta. This brings an extra factor of $1/a^3$ corresponding to the dilution of the created particles.  The frequency
$\tilde\omega_{k}=\omega_{k}/a$ is the rescaled  frequency coming from the energy in conformal time $\frac{1}{2}\tilde \omega_{k}^{2}|\chi|^{2}=\frac{1}{2}\omega_{k}^{2}\frac{|\Psi|^{2}}{a^{2}}$.
Far from the  trapped brane,  once particles have been created in the immediate vicinity of the brane,
\begin{equation}
\rho_\chi\approx \frac{g|\phi-\phi_{1}|}{a^3} \int | \beta_{k}^{f}|^{2} \frac{{\rm{d}}^{3}{k}}{(2\pi)^3}
\end{equation}
This approximation is valid as long as $g\vert \phi-\phi_1\vert \gg \sqrt{\vert 2H^2 - {\cal{K}}^2\vert}$.
When $\xi\ll 1$, the right hand side is at most $\sqrt{2g|\dot \phi|}$, which implies that $ \vert \phi-\phi_1\vert \gg \frac{|\dot \phi|}{g}$, i.e. outside of the interaction region. In the $\xi\gg 1$, the same condition gives $\vert \phi -\phi_1 \vert \gg \frac{H}{g}$ which is also the size of the interaction region.
In the effective description of the inflationary brane motion, this energy density is equivalent to a linear potential and a constant
force towards the trapped brane. In the absence of the inflationary potential, this force may pull  the passing brane  towards the trapped brane.
We are in a position to determine the effective potential in both cases $\xi\ll 1$ and $\xi \gg 1$.

When $\xi\ll 1 $ the potential can be easily calculated as the particle spectrum is gaussian
\begin{equation}
\rho_\chi \approx \frac{1}{(2\pi)^{3}} \frac{a_S^3}{a^3}  (g|\dot \phi_1|)^{3/2} g \vert \phi -\phi_1\vert
\end{equation}
where $a_S\equiv a_1$ and
 the energy density is diluted before  finally tending to zero rapidly.
\\
In the $\xi \gg 1$ regime,
the creation of particles is largely 
dominated
by the tachyonic instability in the vicinity of the trapped brane and depends on
$\int_{t_-}^{t^+} \frac{\Omega_k}{a} {\rm{d}}t$.
The behaviour of this integral is dominated by the region of small momenta.
Let us  concentrate on the low frequency part of the spectrum. In this case the integration region is maximal and the turning points are located at
\begin{equation}
\delta t_\pm \approx \pm \sqrt 2 {\frac{H}{g |\dot\phi_1| }}
\end{equation}
whose norm is always less than $t_1$.
When $K$ is  small, we can expand $\Omega_k$
\begin{equation}
\frac{\Omega_k}{a} \approx \sqrt {2H^2  - g^2 \dot \phi_1^2 \delta t^2}-  \frac{1}{2\sqrt {2H^2  - g^2 \dot \phi_1^2 \delta t^2}}\frac{k^2}{a^2}
\end{equation}
Integrating between the two turning points we find
\begin{equation}
\int_{t_-}^{t^+} \frac{\Omega_k}{a} dt\approx  \frac{\pi}{2} \frac{2H^2-{\cal K}^2_S}{g\vert \dot \phi_1\vert}
\end{equation}
where ${\cal K}_S=k/a_S$ and we have defined
\begin{equation}
\frac{1}{a_S^{2}}= \frac{1}{a_1^{2}} \frac{1}{\pi} \int_{-1}^1 \frac{{\rm{d}}u}{\sqrt{1-u^2}}\frac{1}{(1+ \frac{\sqrt 2}{g\epsilon \sqrt \lambda}u)^{2/\epsilon}}
\end{equation}
where we recall that $g\epsilon\sqrt\lambda \gg 1$.
This result is very important as it shows that 
the spectrum is approximately Gaussian with a width  $\sqrt{g|\dot\phi_{1}|}$ which is much smaller than the band of integration $\Delta{\cal{K}}=\sqrt{2}H$.
Moreover the amplitude of the number of particles depends on $e^{ 2 \pi H^2/g\vert \dot \phi_1\vert}$ which is an exponentially large number. Notice that  the same factor is negligible in the $\xi\ll 1$ regime.
Now that we know that the spectrum is dominated by the tachyonic instability
, we can evaluate the energy density
\begin{equation}
\rho_\chi \approx  \frac{9}{(2\pi)^3} \frac{a_S^3}{a^3} e^{ 2 \pi H^2/g\vert \dot \phi_1\vert}(g\vert \dot\phi_1\vert)^{3/2} g\vert \phi -\phi_1\vert
\end{equation}
Notice that the main difference with the $\xi\ll 1$ is the presence of a large exponential factor coming from the tachyonic instability close to the trapped brane.
Here also, the energy is diluted after going out of the interaction region. The scale factor $a_S$ takes into account the fact that momenta are red-shifted in the integration region

In both cases the effective potential after the interaction region becomes
\begin{equation}
V\approx  m^2\phi^2 + \frac{1}{(2\pi)^3} y(\xi)  \frac{a_{S}^{3}}{a^3} H^3 g \vert \phi -\phi_1\vert
\end{equation}
where the coupling function depends on $\xi$ and reads
\begin{equation}
y(\xi) \approx  {\xi}^{-3/2}
\end{equation}
when $\xi \ll 1$ and
\begin{equation}
y(\xi) \approx  9{\xi}^{-3/2}e^{2\pi \xi}
\end{equation}
when $\xi\gg 1$.
This potential is similar to the ones used in chameleon models \cite{Brax:2004qh}. It has a moving minimum where
\begin{equation}
\phi_{\rm min}= \frac{1}{(2\pi)^3}  y(\xi)  g   \frac{a_{S}^{3}}{a^3} \frac{H^3}{2m^2}
\end{equation}
This minimum goes to the origin as the scale factor increases. This implies that the effect of the trapped brane is only relevant for a few e-foldings after the passage through the interaction region.
Immediately after the passage, the minimum is located at
\begin{equation}
\label{ac1brane}
\frac{\phi_{\rm ini}}{\phi_1} \approx  \frac{1}{16\pi^3} \frac{y(\xi) \epsilon\sqrt \lambda} {3g} \frac{H^2}{M_p^2}
\end{equation}
This gives a criterion for the influence of the minimum of the trapped brane on the motion of the inflationary brane. If $\phi_{\rm ini}\ll \phi_1$, the trapped brane has no influence as the inflaton feels the $m^2\phi^2$ branch of the potential. On the other hand, if $H$ is large enough and $\phi_{\rm ini}\ge \phi_1$, the inflationary brane feels the steep potential due to the trapped brane. In this case, the motion of the inflationary brane is affected for a few e-foldings.

Let us evaluate the jump in the potential at the end of the interaction zone
\begin{equation}
\Delta V \approx \frac{9}{(2\pi)^3}  {\xi}^{-3/2}e^{2\pi \xi} H^4
\end{equation}
when $\xi\gg 1$ and
\begin{equation}
\Delta V \approx  \frac{1}{(2\pi)^3}   \frac{H^4}{\xi^2}
\end{equation}
when $\xi \ll 1$. In both case it depends only on $H^4$ and $\xi$. For large enough $H$, this jump in $V$ can also change the Hubble rate due to the release of energy in the form of radiated particles.\\
As we have considered that  $H$ is  nearly constant and  we have neglected any backreaction on the dynamics of the inflaton while the brane particles are created,
we must impose that $\Delta V / V \le 1$, which gives for $\xi\ll 1$
\begin{equation}
\label{backineq}
\frac{H^2}{M_p^2}\le 3.(2\pi)^3g^2\xi^2
\end{equation}
And for $\xi\gg 1$,
\begin{equation}
\frac{H^2}{M_p^2}\le 3.(2\pi)^3 g^2 \xi^{3/2} e^{-2\pi\xi}
\end{equation}
This condition gives an upper limit on $\xi$.
In fact, if we require that  the Hubble rate $H$ should  be at least of order 1 GeV, $\xi$ must be at most equal to $\xi_{\rm{lim}}\approx 14$ for $g\sim 10^{-1}$. For larger values of $\xi$, the tachyonic instability implies that there is a strong backreaction due to the explosive creation of particules. In this case, the inflaton loses all its energy very quickly and transfers it into radiated particles.

\subsection{ Discussion}

Let us now  make explicit some of the constraints on the parameters of the model.
Using the condition  $\epsilon\ll 1$, we have
\begin{equation}
m \gg \frac{g}{\sqrt \lambda} M_p
\end{equation}
Hence the masses must be large enough and related to the Planck mass.
This implies that
\begin{equation}
\xi \gg \frac{1}{g\sqrt \lambda}
\end{equation}
guaranteeing the constancy of $H$ in the $\xi\ll 1$ region. Using this lower bound we obtain that
\begin{equation}
\label{masse}
m\gg \frac{g^2}{\xi}M_p
\end{equation}
For a fixed string coupling $g_s$, we see that $\xi$ determines the range of masses leading to inflation.

When $\xi$ is large, the mass $m$ can be smaller than the Planck scale.
In the large $\xi$ regime we have
\begin{equation}
g^2\gg \frac{m}{M_p}
\end{equation}
and
\begin{equation}
\lambda \gg \frac{M_p}{m}
\end{equation}
This regime tends to favour small masses and a large compactification radius.

So far we have not tried to connect DBI inflation to observations. Let us now assume that the DBI inflation regime we have analysed is responsible for the phase of inflation resulting in the quantum fluctuations leading to the CMB spectrum . In this case, the COBE normalisation determines the curvature perturbation
\begin{equation}
\label{scal}
{\cal{P}}_s=\frac{H_{\rm{inf}}^2 \gamma}{4\pi^2 M_p^2 \epsilon}
=\frac{g^4\xi^2}{2\pi^2}
=\zeta^2
\sim 10^{-10}
\end{equation}
So
$\zeta\approx g^2 \xi
\sim 10^{-5}$
.
For a reasonable value of the string coupling $g_s \sim 10^{-2}$, we find that inflationary branes whose quantum fluctuations
lead to the CMB anisotropies must be in the $\xi \ll 1$ regime.
We have another observational constraint ; since   gravitational waves have not been detected yet, the ratio $r=\frac{{\cal{P}}_t}{{\cal{P}}_s}$ must be small.
The tensor perturbations spectrum being
\begin{equation}
{\cal{P}}_t=\frac{4H_{\rm{inf}}^2}{\pi M_p^2}
\end{equation}
We deduce
\begin{equation}
\label{ratio}
r=\frac{16\epsilon}{\gamma}\le 1
\end{equation}
With both (\ref{scal}) and (\ref{ratio}), we find
\begin{equation}
\label{upperH}
\frac{H}{M_p}\le 10^{-5}
\end{equation}
We note that this upper bound for the Hubble rate is smaller than the bound given by (\ref{backineq}) and therefore the  backreaction is always negligible in the $\xi\ll 1$ regime.
\\
We have seen that in this regime the slowing down by a trapped brane can be effective if
\begin{equation}
\frac{H^2}{M_p^2} \ge \frac{3.16\pi^3 g\xi^{3/2}}{\epsilon \sqrt \lambda}
\end{equation}
or equivalently
\begin{equation}
\frac{H^2}{M_p^2} \ge {16\sqrt{3}\pi^3}{\xi^{3/2}}\frac{m}{M_p}
\end{equation}
Using (\ref{masse}), we find
\begin{equation}
\label{obscond}
\frac{H^2}{M_p^2} \ge 16\sqrt{3}\pi^3 g {\zeta^{1/2}}
\end{equation}
This is not compatible with (\ref{upperH}).
%
%
So it is not possible to slow the motion of the inflationary brane drastically  after hitting a single trapped brane.
If we assume that there exists  a stack of N closely packed branes in the  interaction region, then equation (\ref{ac1brane}) becomes
\begin{equation}
\frac{\phi_{\rm ini}}{\phi_1} \approx  \frac{1}{16\pi^3} \frac{y(\xi) \epsilon\sqrt \lambda} {3g} N \frac{H^2}{M_p^2}
\end{equation}
And condition (\ref{obscond}) becomes :
\begin{equation}
\frac{H^2}{M_p^2} \ge \frac{1}{N} 16\sqrt{3}\pi^3 g {\zeta^{1/2}}
\end{equation}
Unless we have at least $N\sim 10^9$ branes in the stack, the slowing effects of the stack is not drastic.
The motion of an inflationary brane leading to the CMB spectrum is hardly  affected by trapped branes.
On the contrary, branes in a regime $\xi\gg 1$ are efficiently slowed down
because of radiated particles: a  brane bremsstrahlung.

\section{Conclusion}
We have studied the slowing down of an inflationary brane in the DBI regime along an AdS throat when it hits a trapped brane. We have shown that the brane motion may be slowed down by the creation of particles on the trapped brane. This creation occurs either by parametric resonance or tachyonic instability. The latter case happens when the interaction region is large compared to the Hubble rate. Once the brane has left the interaction region, the effect of the radiated particles is to generate a linear potential whose slope is greatly enhanced in the tachyonic case. This is enough to stop the brane for a few e-foldings until the number of created particles has been diluted.
Branes crossing the trapped brane in a time smaller than a Hubble time, are not drastically affected by the radiation of particles as they are slowed down in manner which does not alter their motion very much. A dramatic effect is only possible when a very large number of trapped branes are stuck in the interaction region.

 An interesting consequence of our results is a selection mechanism for the motion of branes in a throat where trapped branes are also present. Indeed, as the branes go past the trapped branes, the ones leading to a tachyonic instability are successively slowed down. Only the ones with little interaction with the trapped branes can keep on moving. It happens that the ones with small curvature perturbations are the ones which would be little hampered by the presence of trapped branes. This could have an interesting effect selecting the motion of branes whose inflaton would generate small levels of curvature perturbations.

 Another interesting possibilities is the modification of the perturbation spectra as the DBI brane crosses one or many trapped branes. Even if the  motion of the DBI brane is not altered, the slowing down term in the potential having an effective relevance during a short time, one may envisage that the spectrum of primordial fluctuations might be affected by a jump in the derivative of the potential. This may even be the case when the slowing down of the inflaton is negligible. A thorough study of this possibility as well as  models with a jump in the potential energy is left for future work.

\section{Acknowledgements} We would like to thank D. Easson for interesting suggestions and J. Martin for discussions and remarks. This work is partially funded by the "DarkPhys" ANR grant.

\end{document}